\documentclass[iop]{emulateapj}
\usepackage{amsmath}
\usepackage{natbib}

\catcode`_=\active
\newcommand_[1]{\ensuremath{\sb{\mathrm{#1}}}}

\newcommand{\be}[1]{\begin{equation} \label{eq:#1}}
\newcommand{\ee}{\end{equation}}

\newcommand{\pref}{\protect\ref}

\newcommand{\hot}{\ifmmode{8\times10^4~{\rm K}}\else{$8\times10^4$~K}\fi}

\newcommand\lta { \mathrel {\hbox to 0pt {\lower 3.7pt \hbox{$\sim$}
      \hss} \raise 1.7pt \hbox{$<$}}}
\newcommand\gta { \mathrel {\hbox to 0pt {\lower 3.7pt \hbox{$\sim$}
      \hss} \raise 1.7pt \hbox{$>$}}}

\newcommand{\philemail}{judge@ucar.edu}

\newcommand{\val}{
\begin{figure}[] 
\epsscale{0.8}
\plotone{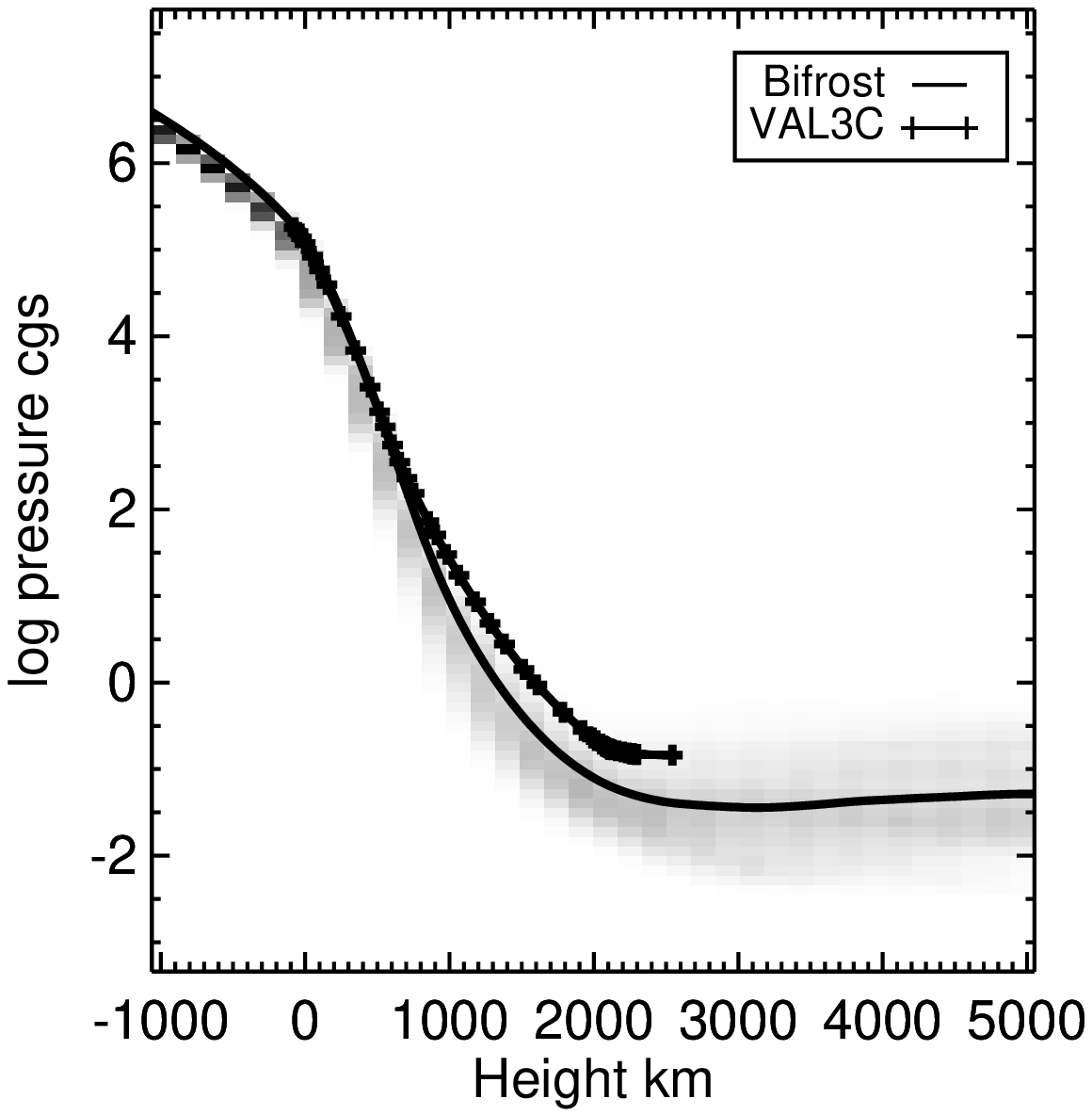}  
\caption{\label{fig:strat} 
The stratification of the VAL3C model is shown together with 
the average and PDF of the stratifications from one random
time step (\#491)  taken from the Bifrost simulation.  
The lower-middle chromosphere is extremely well represented
by the VAL3C stratification up to the point near 2000 km where
dynamical effects start to dominate. Above this height departures from
hydrostatic equilibrium are especially large and have drawn much attention
in the literature. }
\end{figure}
}

\newcommand{\dr}{
\begin{figure}[] 
\epsscale{1}
\plotone{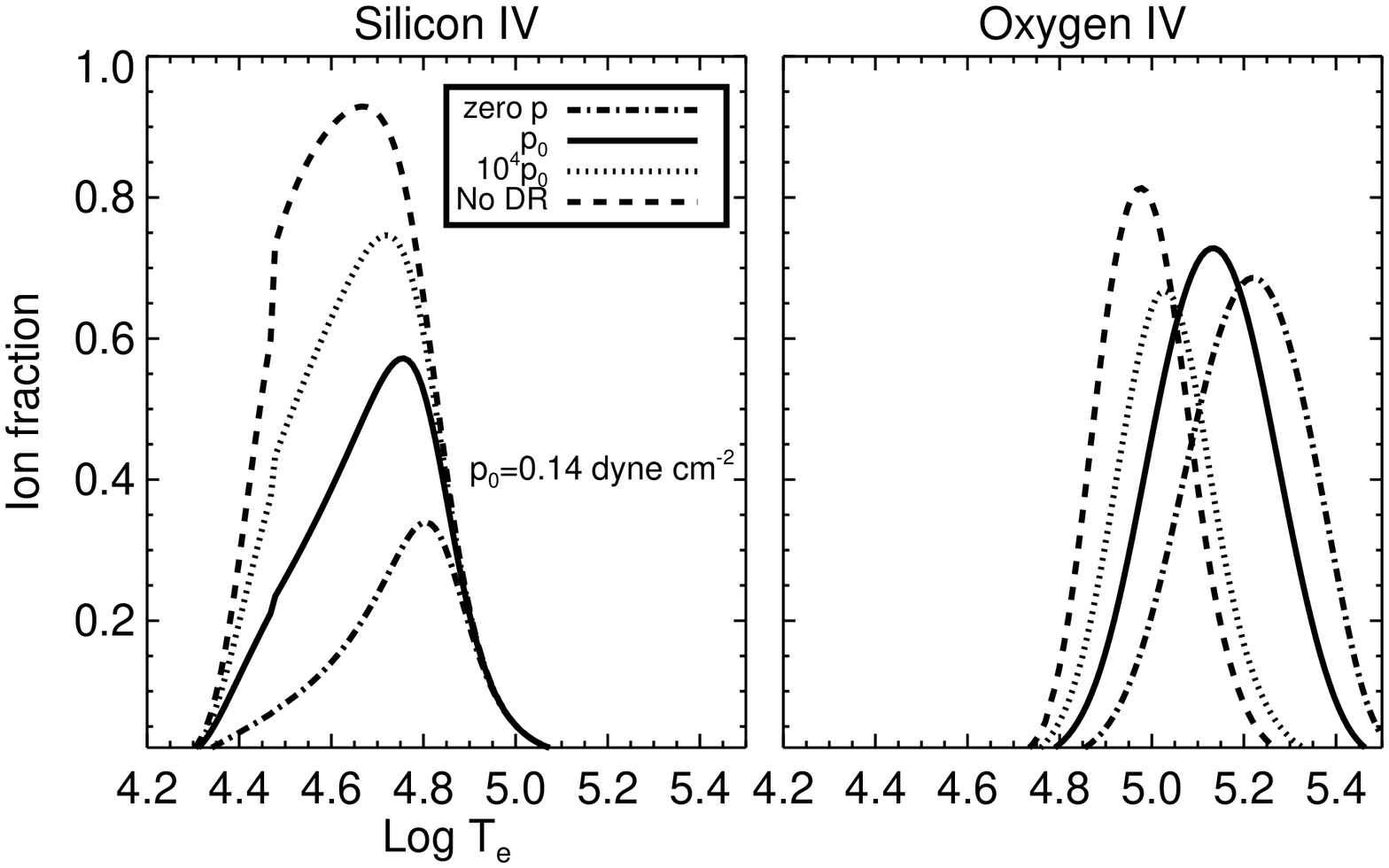}  
\caption{\label{fig:dr} 
The figure compares the ionization balance of Si
and O ions in response to changes in the electron 
density.  In the ``coronal approximation'' (low densities)
the curves do not vary with electron density.  However at 
densities higher than $10^8$ cm$^{-3}$ the effects of metastable
level populations and reduced dielectronic recombination rates 
significantly change, systematically, the relative ion populations. 
Notably, the Na-like \ion{Si}{4} ion's population grows rapidly 
with increasing
density. The opposite is true for the B-like \ion{O}{4} ion. 
 }
\end{figure}
}

\newcommand{\ratio}{
\begin{figure}[] 
\epsscale{1.1}
\plotone{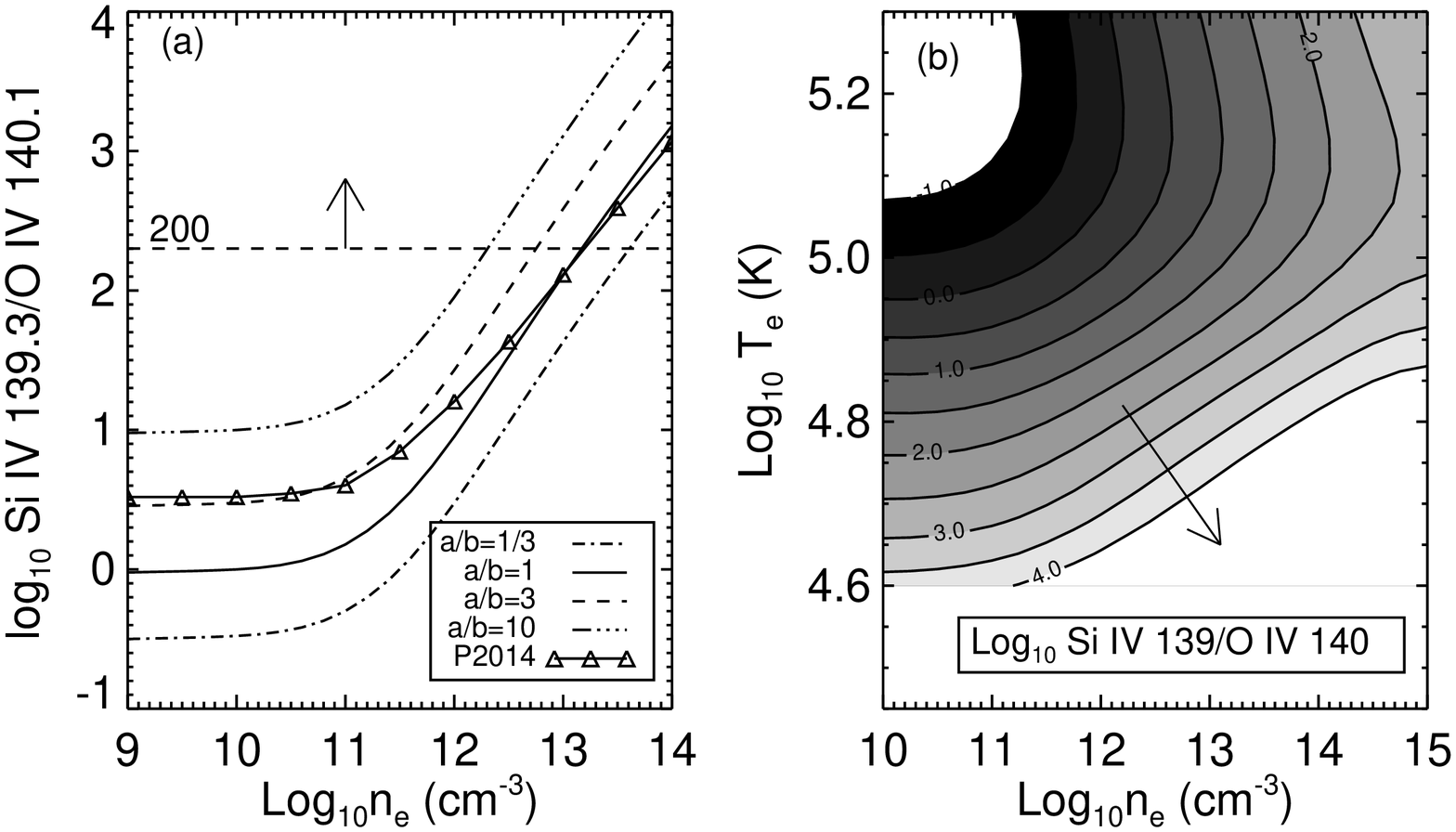}  
\caption{\label{fig:ratio} The computed ratios of line emission is
  plotted as a function of electron density $n_e$ for the lines
  \ion{Si}{4} 139.3 nm to \ion{O}{4} 140.1 nm, assuming constant
  pressure. The densities for the temperature of the \ion{Si}{4} line
  are shown.  The ratios are computed with Equations~\pref{eq:demtwo}
  and \pref{eq:mu}. A ratio of 200, a typical observed value for
  the bombs, is shown as a horizontal line with an arrow.  (b)
  Computed ratios are plotted as a function of electron temperature
  $T_e$ and density $n_e$.  Homogeneous plasmas were assumed.  Lines
  of \ion{Si}{4} amd \ion{O}{4} form at different temperatures so that
  there is a gradient in the ratio $\partial R/\partial T$ below
  $\log_{10} T_e = 5.1$.  }
\end{figure}
}

\newcommand\tableone{
\begin{table}
\label{tab:optical}
\begin{center}
 \caption{Optical depths and escape fractions for model VAL3C}
 \begin{tabular}{lllll}     
 \hline
 \hline
$\tau_{{\rm 140 nm}}$ & $\exp(-\tau_{{\rm 140 nm}})$ & height $z$ 
& $\tau_{{\rm 500 nm}}$ & $p$ \\
& & km & & dyne~cm$^{-2}$ \\
 \hline 
0.3 & 0.74 & 815  & $2\times10^{-5}$& $100$\\
1 & 0.37 & 725  & $3\times10^{-5}$ &200\\
3 & 0.05 & 635  & $5\times10^{-5}$ &400\\
5 & $7\times10^{-3}$ & 590  & $1\times10^{-4}$ &600\\
10 & $5\times10^{-5}$ & 542  & $2\times10^{-4}$ &1000\\
100 & $4\times10^{-44}$ & 450 & $1\times10^{-3}$ &2600\\
 \hline
\end{tabular}
\end{center}
\end{table}}

\newcommand\tabletwo{
\begin{table}
\label{tab:abs}
\begin{center}
 \caption{Absorption lines}
 \begin{tabular}{lllrll}     
 \hline
 \hline
Ion & $\lambda$  & log~gf & $e_{\rm lower}$ &
$p(\tau_0=1)$  \\
& nm  &  & cm$^{-1}$ &
dyn~cm$^{-2}$ &  \\
\hline 
\ion{S}{1} & 139.2589 & -3.35 & 573 & 360\\
           & 140.1515 & -1.18 & 396 & 2.3\\
\ion{Fe}{2} & 133.5409 & -0.93 & 27,620  & 800\\
            & 139.2816$^*$  & -0.483&20,806&60\\
            & 139.3214 & -1.21 & 20,340& 290\\
            & 140.1774 & -0.91 & 22,810& 250\\
            & 140.3101$^*$  & -1.922 & 1,872 & 25\\
            & 140.3255 & -1.25 & 22,810& 550\\
\ion{Ni}{2}& 133.5203$^*$  & -0.19 & 1,507 & 4\\
            & 139.333$^*$   & -0.88 & 0 & 15\\
\hline
\end{tabular}
\end{center}
$^*$Data from Kurucz CD-ROM 23. Other data from NIST's spectroscopic
database.  Partition functions of 50 and 25 were used for \ion{Fe}{2} and 
\ion{Ni}{2} corresponding to a plasma temperature of 6500K. 
\end{table}}

\shortauthors{P. Judge}
\shorttitle{solar UV spectra}

\slugcomment{}

\begin{document}

%
%

\title{UV spectra, bombs, and the solar atmosphere}
\author{  
Philip G. Judge }
\affil{High Altitude Observatory,
       National Center for Atmospheric Research\footnote{The National %
       Center for Atmospheric Research is sponsored by the %
       National Science Foundation},\\
       P.O.~Box 3000, Boulder CO~80307-3000, USA; \philemail}

%
%

\begin{abstract}

A recent analysis of UV data from the Interface Region Imaging
Spectrograph {\em IRIS} reports plasma ``bombs'' with temperatures
near \hot{} within the solar photosphere.  This is a curious result,
firstly because most bomb plasma pressures $p$ (the largest reported
case exceeds $10^3$ dyn~cm$^{-2}$) fall well below photospheric
pressures ($> 7\times10^3$), and secondly, UV
radiation cannot easily escape from the photosphere.  In the present
paper the {\em IRIS} data is independently analyzed.  I find that the
bombs arise from plasma originally at pressures between $\lta80$ and 800
dyne~cm$^{-2}$ before explosion, i.e. between $\lta850$ and 550 km above
$\tau_{500}=1$. This places the phenomenon's origin in the low-mid
chromosphere or above. I suggest that bomb spectra are more compatible
with Alfv\'enic turbulence than with bi-directional reconnection jets. 

\end{abstract}

\keywords{Sun: atmosphere}

\section{Introduction}

\citet[][henceforth P2014]{Peter+others2014} recently 
reported on ``Hot Explosions in the
Cool Atmosphere of the Sun''.  Throughout the paper the \hot{} plasma
is identified as belonging to the solar {\em photosphere}. 
Two major pieces of evidence are presented by P2014 to support
the idea that the bombs originate from photospheric plasma.  P2014
make quantitative estimates of the electron density, leading them 
to conclude
\begin{quote}
``A density analysis based on the IRIS observations of O IV and Si IV
shows that the bombs form in the photosphere''
\end{quote}
The reasoning is that such high density plasma pre-exists only in the
deeper layers.  The second argument concerns the presence of narrow
absorption lines of \ion{Fe}{2} and \ion{Ni}{2} seen against the
bright emission.  Cool plasma exists above the emitting plasma with
sufficient opacity to produce absorption lines. P2014 did not analyze these
lines quantitatively, this is done below. 

In this paper I demonstrate that the $\approx \hot$ radiating plasma
must exist {\em at least} 550 km above the continuum photosphere,
above the canonical temperature minimum region.  I take the essential
observations of P2014 and re-analyze them 
within the following simple picture.  I adopt, with justification, a
1D model of the atmosphere that captures the essential stratification
of 3D dynamical atmospheres to study the transfer of UV continuum and
line radiation, in both the emission (\ion{C}{2}, \ion{Si}{4},
\ion{Mg}{2}) and absorption features (\ion{Fe}{2} and \ion{Ni}{2})
reported by P2014. I assume that electrons are fully thermalized owing
to the high densities encountered.  I re-analyze the physics
determining the ratios of permitted \ion{Si}{4} and spin-forbidden
\ion{O}{4}] lines (``]'' denotes, by convention, a spin-forbidden transition), and the implications of the absorption features.  I
  assume bomb spectra originate from pre-existing plasma that is
  heated to produce the \hot{} plasma, keeping the density
  approximately constant. The latter is assumed to be due to strong
  magnetic forces, to be consistent with the 5 minute lifetimes of the
  bombs that greatly exceed dynamical time scales ($\approx 1000$ km /
  100 km~s$^{-1}$).  Finally I review the analysis of P2014.

\section{Analysis}

Below I use photoionization cross sections from the OPACITY project
\citep{Seaton1987}, other atomic data compiled by \citet{Judge2007a},
IRIS data from P2014 and calibration data from \citet{dePontieu+others2014}.
Photospheric abundances from \citet{Allen1973} are adopted. 

\subsection{Pre-existing stratification and UV opacities}

Under undisturbed conditions, the solar photosphere and chromosphere
are strongly stratified. I therefore discuss bombs in the context of
an average atmospheric model, model  
C (``VAL3C'') of \citet{Vernazza+Avrett+Loeser1981}.  This is a static
1D model. Nevertheless it captures typical density and
pressure stratification of dynamical 3D models.  Much has recently 
been made of
the essential dynamical nature of the chromosphere
\citep[e.g.]{Leenaarts+others2012}, but such dynamical departures
from hydrostatic equilibrium are large only in the tenuous regions
of the upper chromosphere.
Figure~\pref{fig:strat} shows the run of pressure with height $z$ of the
VAL3C model together with a probability distribution function derived
from the 3D Bifrost model ``en024048\_hion'' developed for the IRIS
project\footnote{http://sdc.uio.no/search/simulations}.  It is clear
that VAL3C can be used to study the vertical transport of radiation,
under undisturbed conditions, low in the solar chromosphere.

\val
 
The atmospheres of G-type stars contain enormous opacities from
abundant neutral atoms: Rayleigh scattering from hydrogen and
bound-free transitions from Si, C, Fe, Al \citep[e.g.][]{Allen1973}.
UV radiation cannot escape from or penetrate the undisturbed
photosphere.  Below 152 nm, the opacity in the lower chromosphere is
dominated by bound-free opacity from the ground levels
($3p^2~^3\!P_{0,1,2}$) of neutral silicon.  The \ion{Si}{1}
photoionization cross section is $\sigma_{Si}=6\times10^{-17}$
cm$^{-2}$.  Using initially $n_{Si~I}/n_{Si}\approx 1$ (silicon is
mostly neutral), and an abundance $n_{Si}/n_{H}=3.5\times10^{-5}$, we
find
\be{opacity} k = n_H \frac{n_{Si}}{n_H} \sigma_{Si} \approx 10^{-7}
\frac{n_H}{10^{14} {\rm\ cm^{-3}}} \ \ \ {\rm cm^{-1}}.  \ee 
With these opacities, optical depths $\tau$ and the escape
probabilities $\exp(-\tau)$ for line radiation lying in the \ion{Si}{1}
continuum are listed in Table~1.  Below 600 km, where gas pressures
$p$ exceed 600 dyne~cm$^{-2}$, the optical depth (proportional to
column mass $m$ where $p=mg$ in hydrostatic conditions 
and $g$ is the solar acceleration
due to gravity), fewer than 1 in 100 photons escape from this high
density, opaque plasma, dropping as a double-exponential function
of height.

\tableone

To construct Table~1 I assume unrealistically that bomb radiation does
not affect the plasma surrounding it.  Photoionization of neutrals
reduces UV opacity at 140nm, but this effect is offset partly by
increased recombination as more electrons are made free.

In principle a non-LTE transfer calculation is needed to investigate
the statistical equilibrium of the bomb and its effects on the
surrounding atmosphere.  However, bomb plasma with densities $>
10^{13}$ cm$^{-3}$ in a 1D atmosphere is so geometrically thin ($<0.1$ km) 
and hot that calculations with the RH non-LTE code \citep{Uitenbroek2000} proved
impossible to make converge.

To proceed, two simpler calculations were made. First, RH
was used with downgoing radiative intensities of transition region
lines that equal the brightest observed 
bomb intensities (as shown below), as an upper boundary condition.  
The non-LTE rate equations were
solved for H, C, O, Si, Fe to study the penetration of such bright
radiation downwards and changes in ionization balance and the electron
number densities, keeping the temperature fixed.  
Less than 10\% of the 139.3 nm 
radiation penetrates to or escapes from $z=500$ km.  To understand this
result, I then made the following rough calculations, 
examining the consequence of placing the bright bomb
emission immediately below the temperature minimum region ($z=450$ km,
$p(z)=1600$ dyne~cm$^{-2}$). The undisturbed continuum optical depth
at 140nm is $100$.  The wavelength-integrated intensity of radiation
in bomb B-1 in the \ion{Si}{4} lines is $I_{B-1} \approx
2.6\times10^6$ erg~cm$^{-2}$~s$^{-1}$~sr$^{-1}$.  The photoionization
rate of \ion{Si}{1} by the \ion{Si}{4} bomb emission is
\be{pirate}
P_{1\kappa} = 2\pi \frac{I_{B-1}}{h \nu}
\sigma \approx 70 \ {\rm s^{-1}} 
\ee 
At $z=450$ km the VAL3C electron density is $\approx 4\times10^{11}$ cm$^{-3}$,
and the recombination rate per electron $\alpha \approx 2.5\times
10^{-12}$ cm$^3$s$^{-1}$. Ionization equilibrium gives 
\be{balance} n_{\rm Si~I}/n_{\rm Si~II} = \alpha
n_e / P_{1\kappa} \approx 1/70,
\ee
%
%
i.e. the 140nm continuum optical depth is reduced to 100/70=1.4.  The
observed \ion{C}{2} lines are almost as strong as the \ion{Si}{4}
lines, and H L$\alpha$ will certainly be strong.  But all additional
UV radiation will increase the electron density as
elements become ionized: fully ionized silicon and carbon contribute
an additional $1.4\times10^{11}$ and $10^{12}$ electrons per cm$^3$ at
$z=450$ km. With these additional electrons, 
equation (\pref{eq:balance})
yields $n_{\rm Si~I}/n_{\rm Si~II} \approx 1/16$ with a continuum
optical depth of about 6, stopping all but 2.5\% of the radiation from
emerging.  These calculations explain the RH non-LTE calculations,
so that {\em to conclude, UV bomb emission below 450 km is not
visible to IRIS even in the most extreme case B-1.}

\subsection{Density analysis}

When the radiation in two emission lines depends differently on
electron density $n_e$ and temperature $T_e$, it is obvious that one
line ratio can only broadly constrain possible mean densities in the emitting
plasma.  The general problem of finding electron densities from
thermal but low density plasmas has been discussed by
\citet{Judge+Hubeny+Brown1997}.  Refer to this paper for the notation
used here.  Under standard assumptions each line $\ell$ emits a total
intensity of
\be{demtwo} I_{\ell}= \int_0^\infty \hskip -8pt \int_0^\infty \hskip -8pt \mu(n_e,T_e) H_{\ell}(n_e,T_{e})\ dT_e dn_e \ \ {\rm erg~cm^{-2}~s^{-1}~sr^{-1}, }
\ee 
where $\mu$ is a differential emission measure and $H_{\ell}$ depends on atomic
processes. 
To derive a
mean density $\overline n$ one must specify a form of the source term $\mu$, 
for example in an isothermal plasma at  temperature $T_0$,
$\mu= a \delta(n_e-\overline n) \delta(T_e-T_0)$.  Lines of \ion{Si}{4} and 
\ion{O}{4}] have different 
peak ionization equilibrium
temperature temperatures, $T_{Si}$ and $T_{O}$.
P2014 look for solutions 
for $\overline n$ at constant pressure $\overline n T_e=$constant, so that 
$\mu(\overline n, T_e) \equiv \mu(T_e)$
and where 
\begin{eqnarray} \label{eq:mu}
\mu(\overline n,T_e) &=& \mu(T_e) = a
\delta(T_e-T_{Si}) + b\delta(T_e-T_{O})
\end{eqnarray}
Note that the line ratio yields the mean electron density $\overline
n$ only when the ratio of the emission measures $a$ at $T_{Si}$ and
$b$ at $T_{O}$ is known.  In this regard the calculations of P2014 are
not specified sufficiently to reproduce them.  I find them to be
consistent with $a/b\approx 4$ using calculations with similar
ionization equilibria.  It appears then that they assume the
geometrical thickness of the emitting regions is the same in the Si
and O lines, so that the Si line has roughly 4$\times$ the emission
measure (i.e.  $a/b =4$ in the present notation).

The specific case for \ion{Si}{4}/\ion{O}{4}] intensity ratios has additional
problems:

\begin{enumerate}
\item Si is a low-first ionization potential  (low-FIP) element
  and O is not. Differential abundances $A_{Si}/A_{O}$ between low-
  and high FIP elements are well documented \citep[e.g.][]{Feldman1998a}.
  Si can be enhanced by factors of up to 3 compared to O. 

\item The Si$^{3+}$ and O$^{3+}$ ions have different
  energy thresholds for ionization from the neutral species (58.7 and
  103.7 eV). Consequently 
  ionization equilibrium temperatures differ by a factor of two
  (P2014 use $\log_{10}T_e=$ 4.87 and 5.16).
 
\item Ion populations depend on micro-fields in finite density plasmas
  (above $\approx 10^8$ cm$^{-3}$, \citealp{Summers1974}). Above $10^8$ cm$^{-3}$ 
  collisions reduce rates of dielectronic recombination (DR)
  by perturbing the doubly excited states before the Auger stabilizing
  transition occurs \citep{Summers1974}.  In Na-like Si$^{3+}$,
  DR to Si$^{2+}$ is significantly suppressed
  (see the computations of the
  homologous Li-like ions by \citealp{Doyle+Summers+Bryans2005}).
  Together with increased ionization from metastable levels of B- and
  Be-like ions, at plasma densities above $10^{12}$ cm$^{-3}$ I
  estimate using the approach of \citet{Judge2007a} 
  that the emitted power of Na-like \ion{Si}{4} lines above
  those of \ion{O}{4}] lines by a factor of three to four 
(Figure~\pref{fig:dr}).  

\dr

\item The \ion{Si}{4} line core intensity ratios 139.3/140.2 are 
close to but (at least for bomb B-1 where the ratio is 1.6-1.7)) 
less than 2:1.  Scattering in the outward (observed) 
direction increases the 
139.3 component's integrated intensity \citep{Judge+Pietarila2004}
since the lines are not optically thin, unlike the \ion{O}{4}] lines. 
In essence, downward directed radiation in 
\ion{Si}{4} lines can be scattered outwards, but 
in \ion{O}{4}] lines downward radiation is absorbed by the background
continuum.  

\end{enumerate}

Another problem is that intensities of UV lines of Na-like and Li-like
ions are known to be too weak by factors of several compared with
lines of other sequences.  In the accurate measurements used by
\citep{Judge+others1995}, the Na-like \ion{Si}{4} lines are 2.5 times
brighter than calculations predict. This discrepancy must lie outside
of the high density effects because disk-averaged spectra contain many
spin forbidden lines whose intensities are inconsistent with $n_e \gta
10^{10}$ cm$^{-3}$.  Perhaps this originates in non-equilibrium
ionization \citep{Hansteen1993} or non-thermal distributions of
electrons. Here I simply remind the reader of this additional problem.

Taken together, all these issues indicate that computations of relative
intensities of \ion{Si}{4} and \ion{O}{4}] lines are uncertain to {\em
  at least} a factor of three.  Most importantly, at least four
independent processes can enhance the intensities of the \ion{Si}{4}
line, each by factors of two to three.  When these systematic effects
are compounded, low-density computations may underestimate the ratio
of \ion{Si}{4} to \ion{O}{4}] line intensities by an order of
magnitude, which then would be mistakenly interpreted as an increased
electron density using the above methods.

Figure~\pref{fig:ratio} (a) shows Figure S8 of P2014, generalized to
include different assumed values of $a/b$.  Differences 
with the shape of the curve of P2014 are primarily 
due to the different ionization equilibrium calculations 
used.  I included approximate corrections for the DR suppression following
\citet{Judge2007a}.   
Clearly, {\em the chosen form of
  $\mu(T_e)$ strongly determines the outcome of the density analysis}.
There is no physical reason for choosing particular values of $a/b$ in
the bombs.  When electron heat conduction from overlying coronal
plasma dominates the energy balance, these quantities are physically
related  \citep{Jordan1992}.  But there is no evidence for
coronal emission in the bombs, and there exists a class of solutions
for ``cool loop'' structures invoked to account for long-standing
observational problems in the solar transition region
\citep{Feldman1983,Dowdy+Rabin+Moore1986}.  Such solutions can in
principle have ratios $a/b$ that greatly exceed unity.

Figure~\pref{fig:ratio} (b) shows line ratios computed for homogeneous
plasmas where $\mu(n_e,T_e)=\delta(n_e-n_0,T_e-T_0)$. 
{\em All} regions below the contour marked with the arrow are compatible with
the observed lower limit to the ratio.  When $b \ll
a$ (very little plasma above $T_e=10^5$ K) the line ratios contours 
become more horizontal than vertical.  The ratios are then primarily
sensitive to $T_e$ than $n_e$.  In adopting
equal emission measures for $T=4.87$ and 5.16, P2014 are forced to
look for solutions beyond $10^{13}$ cm$^{-3}$ in their Figure S8 
in order to reconcile observed and
computed line ratios.  Instead, Figure~\pref{fig:ratio} demonstrates that 
the line ratio analysis is subject to an entirely different
interpretation where there is less plasma above $\log_{10} T_e=4.9$. 

\ratio

In conclusion, the mean density is not even defined without
drastic simplifying assumptions, even then they are poorly constrained
and subject to systematic over-estimation by factors of 3 if low
density ionization equilibrium approximations are used.  By
acknowledging possible FIP effects and the anomalous Li- and Na- like
ion intensities, the over-estimate can easily increase to an order of
magnitude or more as the increased Si/O intensity ratio is ascribed
erroneously to pure density increases that decrease the \ion{O}{4}]
emission rate.  

\subsection{Absorption lines}

P2014 report several lines seen weakly in absorption against the bomb
emission.   They find that they are all
slightly blue-shifted owing to flux emergence pushing pre-existing material
upwards, and conclude that ``the absorption features indeed belong
directly to the bombs''.  Here I simply estimate the column mass of material 
needed to account for these weak absorption features.  

\tabletwo

Table~2 lists pressures in the VAL3C atmosphere where the optical
depths in the background atmosphere equal one in those observed
absorption lines for which NIST atomic data are available.  (Optical
depth is proportional to column mass $m$ and $p=mg$). Both \ion{Fe}{2} and 
\ion{S}{1} were assumed to be dominant ionization stages in the 
chromosphere, in making this table.  
The 133.54 nm
line requires the highest optical depths to form a weak absorption
line, the pressure at $\tau_0=1$ being almost 800 dyne~cm$^{-2}$.
This corresponds to a column mass $m$ of 0.03 g~cm$^{-2}$ or a height
of about 560 km.  The line probably requires an optical depth close to
0.1 or even less since it is very shallow (bomb B-1) or absent (other bombs), 
thus these
absorption features must form in plasma originally above about 80
dyne~cm$^{-2}$, near a pre-existing height of 850 km.  
The  absorption of each line is stronger for smaller values of $p(\tau_0=1)$,
The relative 
weakness of the
observed \ion{S}{1} lines implies that \ion{S}{1} is 
mostly ionized above this height, a result confirmed in radiative
transfer calculations with RH. Other than this, the line strengths appear in
qualitative agreement with the absorption line depths seen in the data of P2014
(their Figure 3).

\section{Discussion}

I have summarized and/or demonstrated the following:

\begin{itemize}
\item UV photons from \ion{Si}{4} bombs cannot escape from below 500
  km above the continuum photosphere.
\item Computed \ion{Si}{4} lines have been shown to underestimate
  their brightness by factors of 2-3 in much earlier work, even in
  quiet Sun plasmas.
\item At very high densities claimed by P2014, \ion{Si}{4} becomes yet
  stronger as dielectronic recombination to Si$^{2+}$ is
  suppressed and ionization from metastable levels increases.  Such effects were
  not taken into account by P2014.
\item In at least one case the \ion{Si}{4} lines show evidence of
  a small optical depth.  Scattering of \ion{Si}{4} photons by modestly
  opaque bomb plasma will enhance outward directed intensity by up to
  a factor of 2 \citealp{Judge+Pietarila2004}.
\item Any density analysis of Si/O line ratios hinges on an arbitrary
  assumption concerning the form of the emission measure distribution
  between $\log_{10} T_e=4.8$ and 5.1.  Unfortunately, almost any
  density can be made consistent with the data, given this extra
  degree of freedom.
\item IRIS should observe the permitted multiplet of \ion{O}{4} near 134
  nm if it is to address this problem.

\item Absorption lines with optical depths $\approx 0.1$ form in
normal chromospheric plasma at pressures close to 80 dyne~cm$^{-2}$ 
which originally lies near 850 km. 

\end{itemize}

Let us give P2014 the benefit of the doubt and accept their 
general assumptions in evaluating densities. Even if the assumptions
concerning the form of $\mu(n_e,T_e)$ are accepted, 
their 
calculations miss important physical processes leading them to 
under-estimate the intensities of the
\ion{Si}{4} lines, by a factor of 10 or so.  Consequently they over-estimate
the electron densities by an order of magnitude. 
Even in the most extreme case (B-1), $\log_{10} n_e \gta 12.7$.  This
new constraint 
implies most bomb plasma originated from regions where, before explosion, 
$\rho > 10^{-11}$ g~cm$^{-3}$ 
or a height $< 1180$ km above the undisturbed continuum
photosphere.  This constraint may seem 
uncomfortably above the $850$ km limit derived from absorption lines in bomb B-1. 
But given 
the general difficulties with density diagnosis, and 
the fact that the 850 km limit 
comes from the weakest of the absorption lines visible only
in B-1, 
it seems reasonable
to conclude that, all constraints considered, 
{\em the bombs originate from plasma between $500$ and $\gta 850$ 
km in the pre-bomb
atmosphere}.  If their implicit assumption ($a/b=4$) 
proves incorrect, then the
electron and hydrogen densities will be change accordingly
(Figure~\pref{fig:ratio}(b)). No method is known to 
the present author by which 
this important assumption is testable given these data. New observations
including the permitted 
\ion{O}{4} multiplet near 134 nm should help. 

Are the lower limit pressure and height of 500 km ``photospheric?''
Certainly it lies above the temperature minimum in VAL3C.  However, as
understood in physical terms, the photosphere is the region where the
bulk of the radiative flux is radiated into space.  The solar
photosphere starts therefore where the radial optical depth in the 500
nm continuum equals unity, $\tau_{500}=1$. The upper levels of the
photosphere are less well defined. Allen's (1973, \S{} 77) definition
\nocite{Allen1973} is that the photosphere extends to where
$\tau_{500}=0.005$ which lies 320 km above $\tau_{500}=1$.  Only a
small fraction $\tau_{500}\lta 0.005$ of the Sun's radiative flux can
originate above $\tau_{500}=0.005$, so Allen's definition allows
$>99.5$\% of the solar flux to be radiated.

Does it even matter that the bombs form in ``photospheric'' versus
``chromospheric'' plasma?  Or is this merely semanticism?  Indeed
it matters because of the stratification.  At the high densities of
the photosphere, the Alfv\'en speed is slow for field strengths of a
few hundred Gauss, characteristic of the regions without spots or
pores where these bombs appear. Using Fig~1 of P2014, I use
(optimistically) $B$=400 G.  The Alfv\'en speed $V_A$ at 500 km is 9
km~s$^{-1}$.  Fast reconnection tends to occur at rates with outflows
close to 0.1$V_A$.  For a bomb of say 3Mm size, the reconnection time
scale is of order an hour, an order of magnitude larger than a
typical bomb lifetime.

All things considered, it seems that the bombs arise from plasma
with total particle densities, before explosion, that lie between 550 and 
$>850$ km above the $\tau_{500}=1$. This places the phenomenon's origin in
the low-mid chromosphere or above.

\section{Speculations}

Prompted by the higher Alfv\'en speeds associated with chromospheric
plasma, I speculate on the nature of the bomb plasma.  I also note
that other observational properties of the bombs identified by P2014
are not easy to reconcile with the association of bombs with
reconnection jets.  More generally, reconnection leads to efficient
acceleration of plasma through the change in topology and subsequent
Lorentz force, but the direct heating effects are usually small.  Why
then should we expect to interpret the radiation from strongly heated
plasma such as seen in the IRIS emission lines directly as
reconnection jets?

The 5 minute bomb lifetimes greatly exceed the dynamical crossing
times if we are to accept the proposal that the 100 km/s speeds
represent reconnection jets.  Reconnection must occur over several
minutes in this picture, a perfectly reasonable picture.  However, the
bombs would be expected to exhibit large asymmetries between up- and
down-flowing plasma because the surrounding atmosphere is highly
stratified.  The proposed supersonic jets must surely interact with
surrounding magnetized plasma with observable consequences.  At least
for the overlying chromosphere, no obvious interaction is seen,
instead benign absorption features are observed.  What happens to the
energy of the proposed jets? Perhaps this energy returns to
photospheric layers, guided by closed fields, where it is deposited
but is almost invisible being energetically a small contribution.  I
suggest that a more natural explanation of the broad rather symmetric
lines is Alfv\'enic turbulence trapped in a compact loop-like flux
tube that exists somewhere in the low-middle chromosphere, driven
perhaps by reconnection that occupies a small volume of the system at
any given time.  Bombs are usually associated with magnetic
footpoints of opposite polarities separated by 1000-2000 km,
suggesting such a structure.  This speculation appears worthy of
further study.


\end{document}